\documentclass[runningheads]{llncs}
\usepackage[T1]{fontenc}
\usepackage{graphicx}
\begin{document}
\title{Re/Imagining Smart Home Automation Framework in the era of 6G-Enabled Smart Cities}
%
%
\author{Byungkwan Jung\inst{1}\orcidID{0000-0002-9927-4583} \and
Suman Kumar\inst{2}\orcidID{0000-0002-1375-9724} \and
Adityasinh Manthansinh Chauhan\inst{3}\orcidID{0009-0002-6018-6487}}
\authorrunning{B. Jung et al.}
%
\institute{Troy University, Troy AL 36082, USA \\
\email{\{\inst{1}bjung,\inst{2}skumar,\inst{3}achauhan200456\}@troy.edu}}
\maketitle              

\begin{abstract}
	Smart home automation systems represent a seamless integration of Internet of Things technologies, facilitating the monitoring, management, and regulation of various aspects of our daily life. By leveraging advancements in communication, computing, sensing, and actuator technologies, they hold promises for enhancing the living experience. However, they face challenges such as the need for timely updates, efficient data management, real-time Big data processing, robust security measures, and advanced analytics. In this paper, we propose a novel framework that capitalizes on the capabilities of 6G networks and 6G-enabled cloud computing to address these challenges and improve the overall landscape of smart cities. This framework features enhanced security, data pre-processing, big data intelligence, and security service virtualization in the cloud. Through various application scenarios and a case study—focusing on safe routing during disasters, we demonstrate the utility of this framework and the critical role 6G networks and 6G-enabled cloud computing play in smart home automation.
	
\end{abstract}

\section{Introduction and Motivation}
Modern living has been rapidly evolving within our homes, where an array of electronic devices now seamlessly monitor, manage, and regulate various facets of our daily lives. Advancements in communication, sensing and actuator technologies are transforming these devices into Internet of Things (IoTs) collectively forming a home automation system~\cite{sai2023smart}, enriching the way we experience and interact with our living spaces. From controlling ambiance to ensuring security and efficiency in managing diverse home subsystems like refrigerators, trash cans, and even pet monitoring, these systems have evolved into intelligent, interconnected ecosystems, thanks to high-speed internet connectivity and cloud computing~\cite{ALAA201748}.  Continued advancements in smart home automation systems will undoubtedly unlock endless possibilities for enhancing our homes and ultimately improving our quality of life. However, limitations inherent in IoT devices with constrained storage and processing power, as well as the current state of network itself pose critical obstacles in the future evolution of Smart home automation systems. 

\begin{figure}[h]
  \includegraphics[scale=0.34]{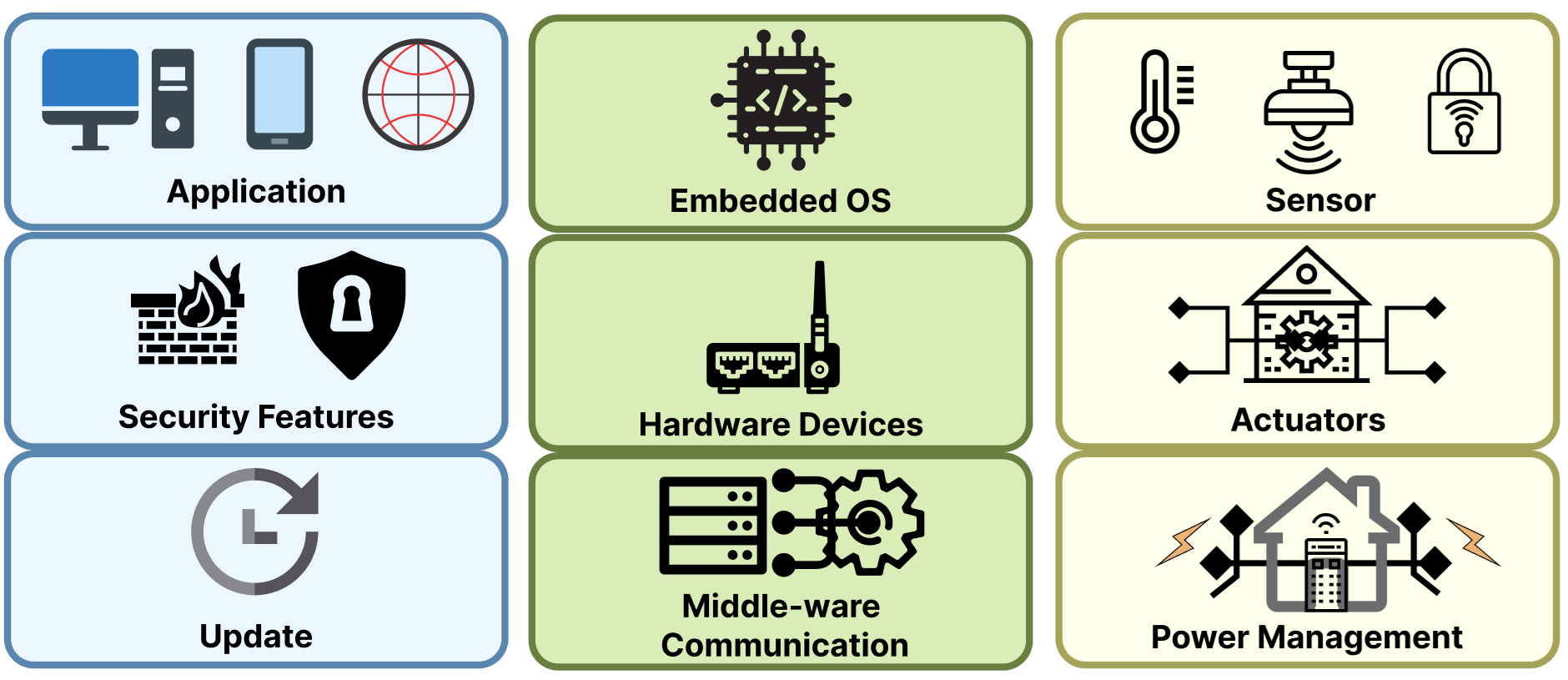}
	\centering
  \caption{State of Art HW/SW Architecture of IoT Devices}
  \label{smartIoT}
\end{figure}
 The architecture of such devices (Figure~\ref{smartIoT}) typically includes IoT applications, security features, and update modules at the top layer. Beneath are embedded OS, hardware drivers, and middleware for communication. Sensors gather data, actuators respond to commands, communication modules enable data exchange, and power management regulates energy usage. However, because of resource limitations this framework struggle with handling big data efficiently, ensuring security against evolving threats, and managing updates effectively~\cite{hategekimana2020iot} \cite{kolias2017ddos}. The dynamic landscape of security and privacy issues surrounding data sharing further complicates the challenge of harnessing collective data intelligence from smart homes, presenting a significant obstacle to realizing a truly responsive smart city ecosystem. These issues become more pronounced for real-time IoT devices which have stricter design constraints, with optimized operating systems, processors, and dedicated sensor interfaces.  They often keep services on the device itself, as network latency becomes a critical bottleneck even with low-latency wireless protocols such as Zigbee and Z-wave. Furthermore, the outer network typically relies on legacy technologies like 4G and 5G, which suffer from latency and connection reliability issues. 

There have been significant efforts to accommodate the growing number of IoT devices and their demanding applications \cite{wang2022sustainable}. The deployment of fifth-generation (5G) cellular technology has significantly enhanced network capacity and reduced latency compared to fourth-generation (4G) networks, leading to a surge in adoption of wireless devices. However, with the continuous influx of IoT devices into the network, the demand for capabilities such as massive data rates, ultra-low latency, enhanced computation power, scalability, and heightened security measures continues to escalate. The upcoming sixth-generation (6G) technology aims to tackle these evolving demands through innovative concepts like swarm networks, self-sustained networks, and edge intelligence. With the integration of 6G-enabled cloud computing, the service burden will shift from application devices to virtualized infrastructure, offering greater flexibility and efficiency. Technologies such as big data intelligence, digital twins, and the ever expanding landscape of cloud-based services are reshaping the way we perceive, receive, and consume services. In this paper, we demonstrate how a 6G-enabled cloud coupled with 6G network effectively addresses the numerous challenges currently affecting smart home automation systems.

In this paper, a smart home framework that harnesses the capabilities of the 6G network and 6G-enabled cloud is presented. The framework enhances the spectrum of services offered to smart homes and smart city and also offloads as many tasks as possible to the cloud from IoT devices, thereby enabling the devices to perform minimal tasks. we demonstrate the framework through a case study: smart home automation systems computing real time safe routing in disaster situations. The proposed framework consisting of 5-layer service and infrastructure architecture for the cloud and a 4-layer IoT architecture for smart home side communicating over 6G networks includes the following features:

\begin{itemize}
    \item Enhanced zero-day security and device auto-updates.
    \item Data pre-processing and advanced security services in the cloud utilizing the ultra low latency and high reliability features of 6G networks.
    \item Big data intelligence and analytics for secure and optimized operations.
    \item Dynamic management of security compliance landscape.
    \item IoT security services virtualizations in the cloud for efficient, flexible, and scalable secure smart home operations.
\end{itemize}

The paper is organized as follows: Section \ref{Related} presents background, challenges, opportunities and a brief past work. Our proposed novel smart home automation framework is presented in Section~\ref{System}. Section \ref{experiment} describes application scenarios and a case study illustrating  safe routing in the even of disaster within our proposed framework. The paper concludes with possible future work in Section \ref{conclusion}.

\section{Background, Challenges, Opportunities, and Past Work}
\label{Related}

\subsection{Smart Home Automation}
\begin{figure}[h]
  \includegraphics[scale=0.26]{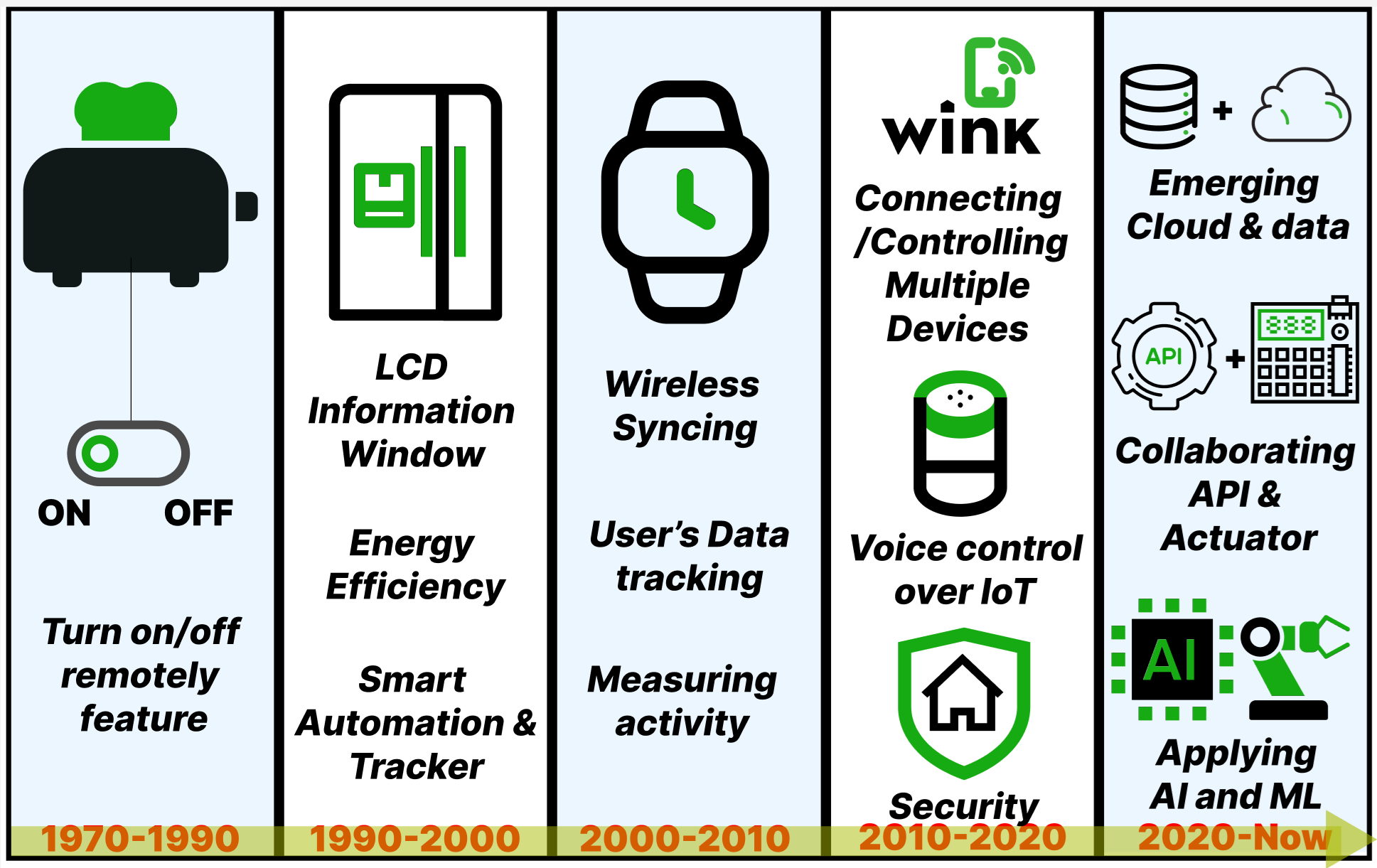}
	\centering
  \caption{Home Automation Evolution}
  \label{HomeAutoEvol}
\end{figure}

\subsubsection{Evolution}

Figure~\ref{HomeAutoEvol} shows the evolution of Home automation system. In more recent decade, say 2000s, with the advent of wireless tech, basic control features like automated air conditioning and lighting emerged. However, not all devices in a home are managed by one interface yet. it is still isolated but connecting multiple home appliances \cite{sovacool2020smart}. As 4G and 5G technologies are introduced and IoTs are becomes a new trend, not only the home appliances but also other devices becomes a part of the devices managed by interfaces. With higher data rates and faster responses from low latency, more data throughput and diverse activities are in place. Home systems supporting voice over IP (VoIP), home security sensors, and smart lights and thermostats that can be managed by a individual remotely becomes common in many home system today. However, security and privacy concerns loom large, necessitating rigorous device interaction analysis for user safety \cite{mladenova2023smart}. Future smart homes, leveraging 6G, promise intelligent management and decision support via localized cloud services (edge intelligence), advancing towards all-in-one applications \cite{antzoulis2022iot}.

\subsubsection{State of Art and Challenges}
\paragraph {Update Mechanism:}
Smart home IoT devices often suffer from outdated software components, necessitating frequent and timely updates to mitigate vulnerabilities \cite{prakash2022software}. Many smart home IoT devices still rely on outdated and vulnerable software components, posing security risks. 

\paragraph {Data Management:}
IoT systems in smart homes generate vast amounts of data crucial for improving customer experience and system monitoring  \cite{vajagic2022smart}. The influx of data from IoT devices can overwhelm traditional data management systems, necessitating efficient data comparison, storage optimization, and processing techniques.

\paragraph {Real Time Data Processing and Decision Making:}
Authenticated data transfer methods facilitate real-time data processing between IoT devices and the cloud, enhancing smart home IoT capabilities \cite{condon2022design}.  However, large volumes of real-time data can lead to bottlenecks and delays, hindering timely decision-making and system responsiveness.

\paragraph {Security Measures:} 
Securing cloud-based IoT involves implementing encryption, firewalls, secure communication protocols, multi-factor authentication, and secure data storage techniques \cite{mishra2023design}. These remain major concerns in smart home automation systems, with potential vulnerabilities in communication protocols, firmware updates, and network infrastructure. Ensuring the privacy and security of data transmitted and stored by IoT devices remains a significant challenge in smart home environments.

\paragraph{Big Data Analytics, Prediction and Modeling:}
IoT devices play a crucial role in leveraging machine learning (ML) and IoT technologies for various applications. Data driven security mechanisms such as Intrusion Detection Systems (IDS) analyze patterns of attacks and mitigate security threats in real-time \cite{sasirekha2023synthesis}. Integrating diverse data sources, protocols, and formats requires interoperability standards and middleware solutions.

\subsection {6G-Network}
Increase in number of IoTs in the network and personal data usage throughout 5G deployment exposed its limitations. 6G holds the promise to achieve high bandwidth and ultra low latency of the order of fraction of a millisecond \cite{you2021towards}. 6G aims to utilize higher frequency wave to significantly increase the data rates. With estimated bandwidth upto 1 Tbps, real-time decision making services, such as safe passage proposal and real-time incident detection and monitoring can benefit from the huge data rates.
 Real-time data analysis and distribution using machine learning and artificial intelligence are key services which can mutually benefit one another \cite{sun2020machine}. Techniques like evolutionary computing, neural computing and fuzzy systems can enhance the performance of 6G mobile networks by effectively managing massive data loads and diverse scenarios~\cite{guo2021enabling}. 6G-enabled IoT systems can leverage Federated learning(FL) among IoT devices and edge computing to ensure trust and low energy \cite{adhikari20226g}. Data leakage from data training process increase concerns as not all IoTs are equipped with computing power to perform cryptographic algorithms \cite{you2021towards}. IoT devices, having feeble computation power and limited functionality unlike desktops, are targeted as a point of attack in the connected network. Due to heterogeneous characteristic of IoTs, the 6G network  aims to incorporate diverse security capabilities.

\subsection{6G-Enabled Cloud}
As the number of IoT devices increases, centralized cloud servers struggle to meet the demands of low latency high throughput applications \cite{mao2023security}, motivating the need for edge computing and a distributed cloud framework \cite{garcia2015edge}.  In a 6G network environment, end systems can deliver intelligence \cite{giordani2020toward}, by moving heavy computation services such as big data analysis close to data sources \cite{adhikari20226g}. Big data services, such as cloud-based IoT healthcare networks, utilize data from body sensors and apply machine learning \cite{mehta2023big}. IoT devices can leverage machine learning and AI technologies for real-time monitoring, and can be deployed on cloud platforms enhancing efficiency and accuracy in smart home environments. Security measures in cloud-based IoT networks, such as fully homomorphic encryption aided by semi-trusted servers \cite{rezaeibagha2021toward} and CP-ABE \cite{hahn2020efficient}, employ encryption/decryption on attributes/data generated by IoT devices. Intrusion Detection Systems (IDS) analyze possible intrusions based on previous attack patterns for smart home systems \cite{sasirekha2023synthesis}. Through the utilization of NFV and SDN, a myriad of IoT virtualization techniques have been proposed \cite{alam2020survey}. These techniques are limited in compatibility and security. 

\begin{figure*}[h]
  \includegraphics[scale=0.18]{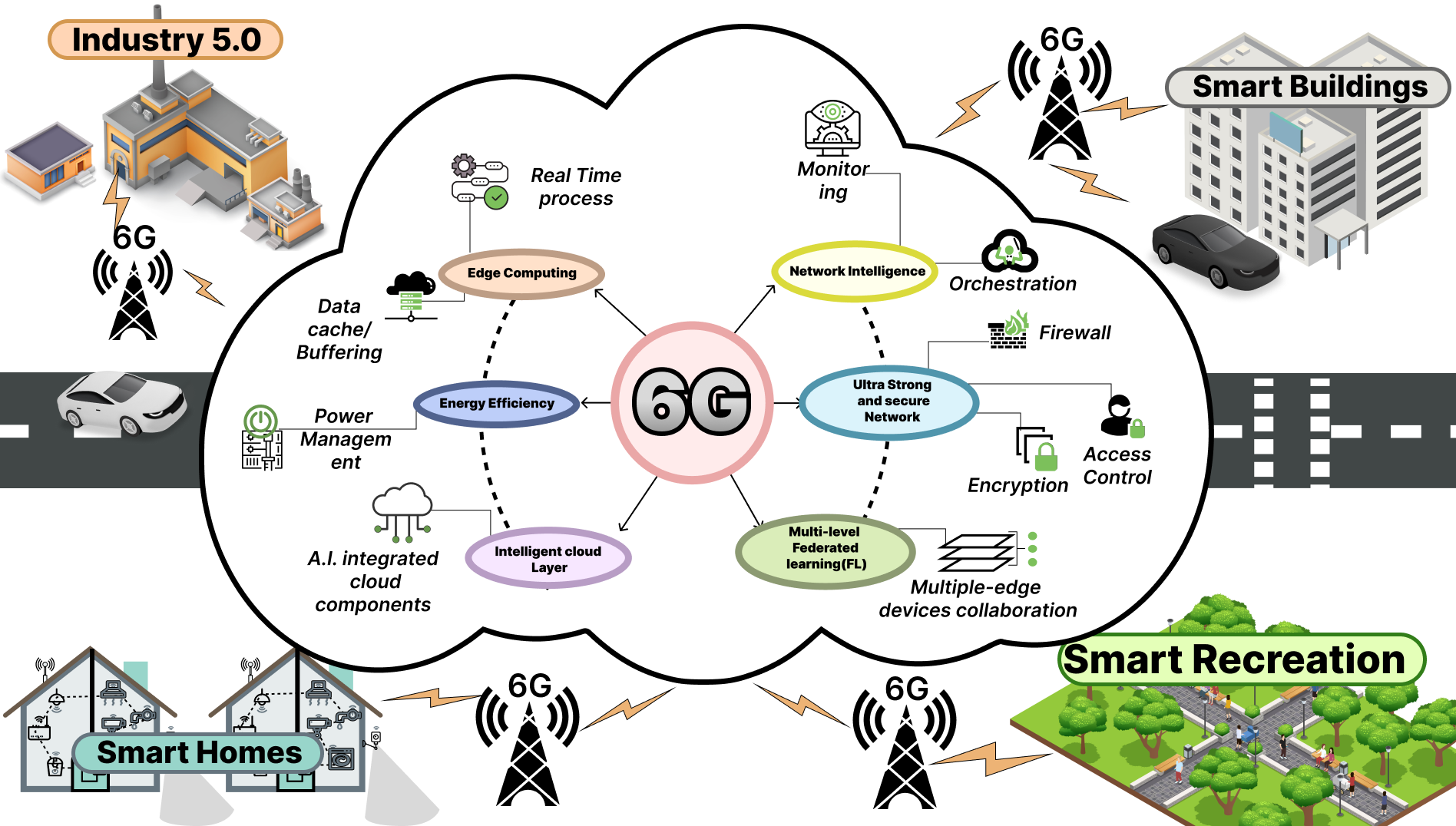}
	\centering
  \caption{6G Enabled Smart City Eco System}
  \label{6GSmartCity}
\end{figure*}

\subsection{Past work on Home Automation Systems}

Various methods, such as analyzing user agent strings using IoT inspector's dataset and employing OTA smart updates, ensure uninterrupted service through smart patching \cite{prakash2022software,srinivas2022deployment}. Effective data management techniques, including data comparison, storage optimization, and utilization of big data services, optimize resource usage and ensure security with minimal delays \cite{mehta2023big}. Comparative analysis of IoT cloud providers aids in selecting the most suitable platform for industrial and home automation applications based on various factors like latency and user-friendliness \cite{haghnegahdar2022iot}. Advanced encryption schemes, decentralized blockchain-based security solutions, and hardware-based isolation mechanisms are proposed to secure and protect IoT systems \cite{hahn2020efficient,rezaeibagha2021toward}. An authenticated search method for data transfers between IoT devices and the cloud is proposed  for real-time data processing in IoT applications \cite{condon2022design}. To assess Data sharing in Cloud-Assisted IoTs, five IoT cloud providers (Adafruit IO, Amazon Web Service, Blynk, Thingspeak, and Ubidots) are compared for industrial and home automation where latency, interval for update, user-friendliness, IFTTT compatibility \cite{haghnegahdar2022iot}.  


\section{Smart Home Automation Architecture}
\label{System}

Figure~\ref{6GSmartCity} shows a 6G-enabled smart city infrastructure that serves as the underlying structure by integrating diverse network applications and their associated services, facilitating seamless data communication across the smart city's nodes. Leveraging its advanced features and enhanced network capabilities, 6G technology offers substantial operational flexibility for smart city infrastructures. Noteworthy attributes include an expanded spectrum, ultra-low latency, guaranteed Quality of Service (QoS), integrated intelligence, built-in optimization capabilities, broader integration capabilities, an improved air interface, and reduced operational costs all the while providing ultra-high reliable connectivity. These features are poised to transcend barriers such as time and space constraints, enabling efficient communication among various smart city devices and systems \cite{you2021towards}. By co-designing communication and management protocols, smart cities can achieve cost reductions, enhance data transmission rates, and foster the proliferation of innovative applications, particularly in areas such as industry, transportation, recreation, and public safety. Ultimately, the integration of 6G technology promises to revolutionize the connectivity landscape of smart cities, driving efficiency, sustainability, and improved quality of life for residents.
\begin{figure*}[!htp]
  \includegraphics[scale=0.18]{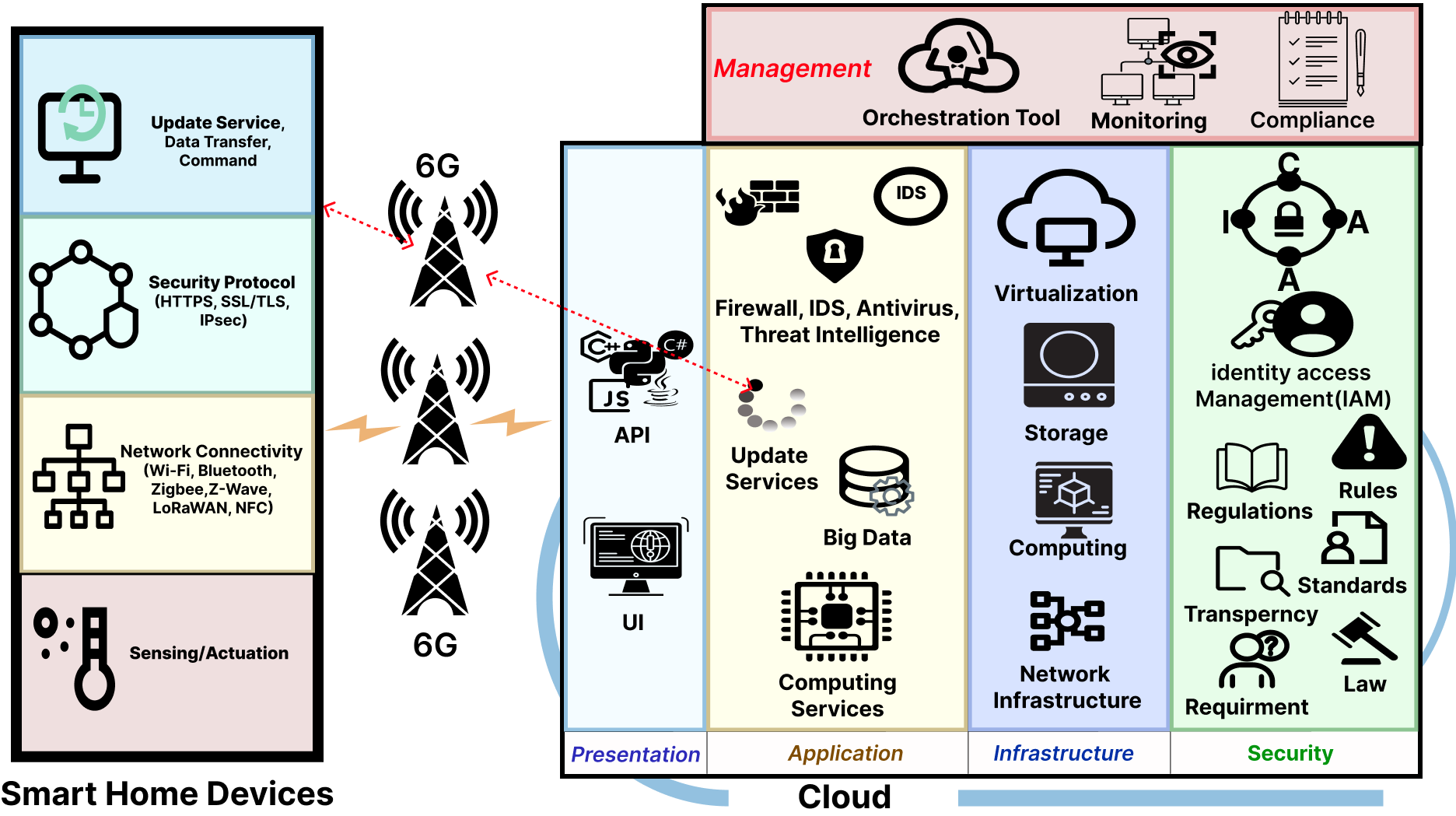}
	\centering
  \caption{Proposed Layered Smart Home Automation Framework}
  \label{SHAFrame}
\end{figure*}

Energy consumption is a significant concern when transmitting raw data from home units to the cloud \cite{condon2022design}. However, recent advancements in energy technologies, ranging from renewable sources to wireless energy transfer mechanisms, offer promising solutions. With these advancements, it's assumed that home devices will consistently have access to power. Given their limited computational and storage capabilities, it is more efficient for these devices to process raw data in the cloud. The cloud possesses sophisticated techniques to address various data issues, including bad data, missing data, anomalies, and outliers, ensuring accurate and reliable data processing. Privacy and security issues can be addressed by using Federated learning~\cite{Duan2023} from each home. Using 6G-enabled edge intelligence, sophisticated machine learning algorithms and artificial intelligence models can be deployed. Raw sensor data can effectively processed for a diverse range of applications, from real-time monitoring to  analytics.

In the context of a 6G-enabled cloud, end systems have the capability to provide intelligence by offloading heavy computational tasks, such as big data analysis, to the network's edge where the data is located. This approach enables edge intelligence to significantly reduce latency by handling data storage, processing, and analysis closer to the data source  \cite{giordani2020toward}, \cite{adhikari20226g}. Moreover, edge intelligence plays a crucial role in enhancing privacy by shifting computational tasks to the network edge, sensitive data can be processed locally, minimizing the need for data transmission and storage in potentially vulnerable centralized locations. Furthermore, advancements in generative AI enable the automatic generation of updates based on the detection and identification of vulnerabilities. These updates can be rapidly tested and deployed to end devices, ensuring timely mitigation of security risks~\cite{10198233}. 

 
Overall, the integration of edge intelligence in 6G networks promises to revolutionize data processing, privacy protection, and security measures, paving the way for a more efficient and secure communication ecosystem. Our proposed framework, leveraging the described 6G features, is illustrated in Figure~\ref{SHAFrame}. Next, we describe our proposed framework with layered architecture for both terminal home devices within the home automation system (SHA) and a layered cloud framework, comprising Security, Infrastructure, Application, Presentation, and Management layers.

\subsubsection{Layered Service architecture in Cloud}

\paragraph {Security:} The security layer includes critical tasks essential for safeguarding data and resources within the system. It is responsible for establishing a robust security framework that instills trust and confidence among stakeholders. It hosts Identity and Access Management (IAM), a pivotal component responsible for managing user identities, home IDs, facilitating access control. Encryption and Key Management form integral parts of this layer, playing crucial roles in protecting data both at rest and in transit. Key management module of this layer is responsible for generation, distribution, and rotation of encryption keys, ensuring the confidentiality and integrity of encrypted data \cite{mao2023security}. The changing landscape of security rules and government regulations sit at this layer, serving as the repository of current security standards and protocols. Any additional security requirements related to data movement, whether internal or external to the cloud environment, are addressed within this layer.

\paragraph {Infrastructure:} 
 The infrastructure layer serves as the bedrock for both physical and virtual infrastructure within the system. Within this layer, the networking infrastructure plays a central role in facilitating communication among various physical and virtual network components, such as routers and switches. These components form the backbone of the system's connectivity, enabling seamless data exchange and resource utilization. Resource virtualization ensures scalability, flexibility, and efficiency in resource allocation and utilization \cite{ogawa2019iot}. The infrastructure layer relies on security features provided by the layer below to enforce access control and security measures. Conversely, the security layer depends on access to the infrastructure layer for key management and storage of security rules, establishing a symbiotic relationship.

\paragraph {Application:} 
The application situated above the security layer contains application logic and services. Within this layer, security measures encompass firewalls, intrusion detection systems (IDS), antivirus software, and threat intelligence services, working harmoniously to ensure the security of both cloud infrastructure and IoT devices \cite{sasirekha2023synthesis}. The application layer contains data preprocessing services tasked with cleaning and refining raw streaming data originating from SHA devices. These services deal with issues such as missing data, anomalies, and erroneous entries while performing tasks such as aggregation and filtering to optimize the data. Computational services are dedicated to executing intricate algorithms for analytics, machine learning, and automation. These services empower the system to derive actionable insights from the data, driving informed decision-making and facilitating automated processes. Update service is a critical component that interfaces with terminal SHA devices to identify and rectify any vulnerabilities promptly. By managing and implementing updates on demand, this service ensures that the devices remain current with the latest security patches and software updates. The application layer hosts big data services, facilitating comprehensive city-wide data analysis. This capability not only supports informed decision-making but also enables privacy-preserving federated learning, bolstering data security and privacy \cite{vajagic2022smart}. The application layer is positioned above the security layer to enforce adherence to the requisite security standards and protocols, fostering a secure and reliable computing environment.

\paragraph {Presentation:} 
The presentation layer encompasses two vital components: user interfaces (UI) and Application Programming Interfaces (APIs). UIs serve an array of dashboards and handheld device interfaces. These interfaces empower users to seamlessly interact with their home devices, enabling them to observe, monitor, and issue commands with ease. From adjusting settings to monitoring activity, UIs provide intuitive pathways for users to engage with their smart home ecosystem. On the other hand, API play aims to extend the functionality of the system by providing programmable interfaces to third-party applications and developers \cite{mladenova2023smart}. These APIs offer controlled access to data and devices, enabling developers to build cloud-based services and applications tailored to specific needs. Additionally, APIs grant access to a variety of services located within the Application layer, fostering interoperability and enabling the integration of diverse applications into the smart home ecosystem.

\paragraph{Management:} 
The Management Layer is and umbrella layer, responsible for overseeing, orchestrating, and automating system operations. Its tasks are enforcing security policies to ensure compliance with regulations, standards, and policies \cite{prakash2022software}. Additionally, the Management Layer handles resource provisioning, automatic deployment, and system scaling through monitoring. By generating a range of metrics, it offers administrators visibility into various aspects of the system, including compliance status and system events, facilitating effective tracking and management of tasks.

\subsubsection{Layered Framework for Terminal SHA Devices}
The SHA devices are streamlined to bare minimum essential functionalities only, enabling them to perform key tasks such as data transmission, updates, command reception, and secure communications. These devices operate on a four-layer stack, prioritizing communication capabilities as outlined below.
\paragraph{Update, Data transfer and Command:}
At this layer update services are responsible for tracking, scheduling, and performing updates. The update function is in communication with the update service at the cloud \cite{prakash2022software}. Cloud counterpart can also perform a rollback on the updates. In addition to these, this layer is responsible for raw data transfer using the massive data rate, ultra-low latency, and highly reliable feature of 6G. The raw data is preprocessed in the cloud as described in the last subsection. In addition, SHA devices can receive commands to perform suitable tasks.

\paragraph{Security Layer:}
The Security Layer is critical as it ensures the protection of data and communications within the framework. It encompasses encryption, authentication, and access control mechanisms, which are continually monitored and updated through the cloud's update feature.

\paragraph{Network:}
This layer serves as the backbone for network connectivity, offering a comprehensive suite of data communication protocols \cite{garcia2015edge}. It includes various layer 2 protocols like Wi-Fi, Bluetooth, Zigbee, Z-Wave, alongside addressing and reliable transport protocols

\paragraph{Sensing and Actuation:}
This layer orchestrates the operation of sensors for data acquisition and actuators for executing desired and/or necessary actions \cite{sai2023smart}.

\section{Application Scenarios and a Case Study}
\label{experiment}

\begin{figure*}[h]
  \includegraphics[scale=0.18]{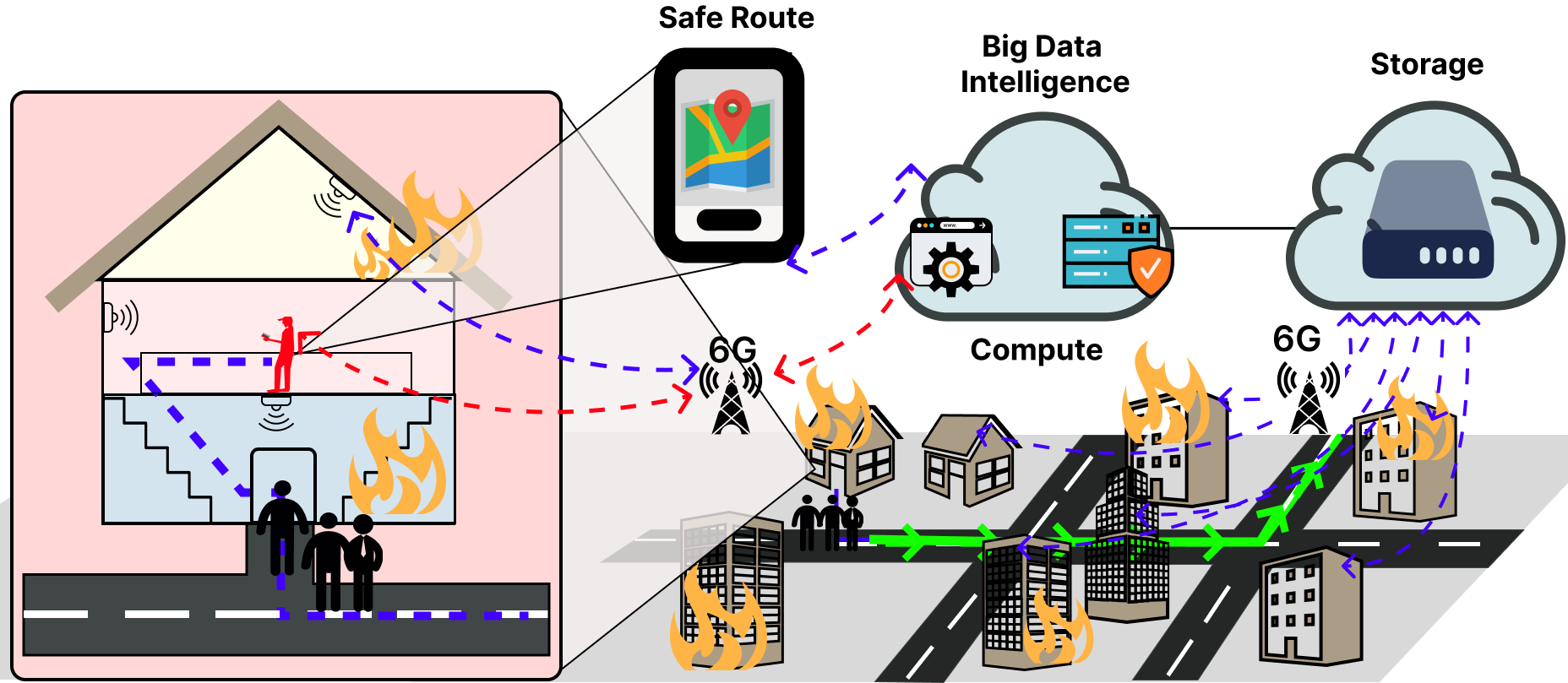}
	\centering
  \caption{Smart Home Automation Framework: Safe Route Computation and Dissemination in the Event of Local Disaster (Fire) }
  \label{SafeRouting}
\end{figure*}


As cities attract more residents seeking better opportunities and lifestyles, the demand for essential services rises, leading to increased living costs, crime rates, and strain on infrastructure. The proposed architecture utilizes  the data collected from smart homes in providing critical information and insights to city government and residents. This section delves deeper into application scenarios and a case study  within smart cities operating in a 6G network environment, where our proposed smart home automation architecture yields significant benefits as compared to existing ones. 
\subsection{Application Scenarios}
The proposed framework enhances public safety and aids in search and rescue efforts for missing individuals. Sensors in nearby buildings swiftly detect suspects, triggering immediate alerts to local authorities, thus preventing their escape. Continuous monitoring and seamless data sharing among buildings enable real-time suspect tracking. Surveillance cameras monitor civilian activities, promptly identifying and reporting suspicious behavior. Anomalies are quickly communicated to the smart city cloud, which collaborates with neighboring cities for regional monitoring and pursuit. Leveraging 6G network services, IoT devices access cloud resources for functionalities like facial recognition and object identification. Citywide  home intelligence emerges as a prominent application of the proposed architecture, capable of performing citywide analyses and pattern recognition based on data collected from multiple IoT devices integrated into smart home automation systems. With a diverse array of sensors deployed both inside and outside buildings, civilians can receive personalized analysis reports detailing various aspects of home activities. 

\subsection{Real Time Safe Routing in the Event of Disasters}

The proposed framework offers a comprehensive solution for enhancing safety and rescue operations in both home and city. We present a case study that illustrate how the proposed framework proves immensely beneficial in real-time applications, particularly in the event of natural disasters. Figure~\ref{SafeRouting} illustrates a scenario in a city where multiple buildings are engulfed in fire. Until emergency services arrive on the scene, civilians must take immediate action to avoid potential danger zones. Without accurate situational details, choosing an improper evacuation route could result in casualties. In a smart city environment, 6G technology enables real-time dissemination of situation updates to all individuals through their personal devices. Real-time data analysis assesses nearby conditions, while sensors both inside and outside buildings continuously monitor the spread of the fire, guiding people to safety. This case study demonstrates how the proposed framework not only directs civilians inside buildings but also guides them away from the fire danger zone.

\paragraph{Smart Home Sensing and data collection:} 
Fire and smoke sensors deployed both inside and outside buildings serve as crucial data sources within the home automation systems. The architecture integrates three main modules: sensing and communication, databases for storing dynamic and historical data on fire spread and rescue operations, and a compute engine for data analysis. These data are stored in two types of databases: static and dynamic. Static data include information such as building layouts and road maps, while dynamic data provide real-time updates on fire and environmental factors influencing its spread. Ultra-low latency and massive capacity of the 6G network swiftly transport  real-time raw data to the abundant compute and storage resources of the cloud.

\paragraph { Edge Cloud Compute Engine:}
In the framework, the 6G-enabled cloud brings computing capabilities to the edge of the network. Here, the compute engine utilizes machine learning models like Bayesian networks (See~\cite{liu2017safernet}) to assess safety probabilities for various route segments. During a disaster, the cloud can prioritize this computation intelligently. By analyzing historical data alongside real-time inputs, the engine discerns the safest routes for fire rescue operations. Continuously updating with streaming real-time data from fire sensors, the Bayesian network ensures precise and adaptive decision-making in dynamic fire scenarios.

\paragraph{Safety Route dissemination:}
The proposed home automation architecture facilitates coordination and decision-making  in fire rescue operations by leveraging real-time data from smart homes and city infrastructure. Through integration with 6G networks and cloud computing, the architecture empowers civilians to make informed decisions about safe route options. This information is delivered directly to their mobile devices and on the map application.

\paragraph{End-to-End Communication Between Civilians in Distress and Fire Authorities:} 

Information regarding trapped civilians, collected by a variety of home automation sensors (such as indoor localization and user behavior monitoring without privacy violations), is relayed to emergency services, ensuring they have timely and accurate information about those in need of rescue.

\section{Conclusion \& Future Work}
\label{conclusion}



An innovative approach to address the challenges  of state-of-the-art  smart home automation systems is proposed. By harnessing the capabilities of 6G networks and 6G-enabled cloud computing, the proposed architecture offers novel solutions to enhance security, data processing, and analytics in smart cities, improving the efficiency and functionality of smart cities. It features enhanced security measures, data preprocessing, big data intelligence, and IoT service virtualization in the cloud, contributing to the development of a robust and efficient smart home ecosystem and paving the way for the realization of true smart cities. Through explorations of various application scenarios and a case study, the impact of the proposed architecture is demonstrated. The future work could focus on further refining the architecture and exploring additional application domains. This could involve optimizing security measures, enhancing data processing algorithms, and expanding the scope of IoT services virtualization in the cloud. Additionally, exploring novel ways to integrate emerging technologies such as AI and edge computing could unlock new opportunities for advancing smart home automation systems in the era of 6G-enabled smart cities. 

%
%
%
\bibliographystyle{splncs04}
\bibliography{bibfile_2024SP}

@article{sovacool2020smart,
  title={Smart home technologies in Europe: A critical review of concepts, benefits, risks and policies},
  author={Sovacool, Benjamin K and Del Rio, Dylan D Furszyfer},
  journal={Renewable and sustainable energy reviews},
  volume={120},
  pages={109663},
  year={2020},
  publisher={Elsevier}
}

@article{haghnegahdar2022iot,
  title={From IoT-based cloud manufacturing approach to intelligent additive manufacturing: Industrial Internet of Things—An overview},
  author={Haghnegahdar, Lida and Joshi, Sameehan S and Dahotre, Narendra B},
  journal={The International Journal of Advanced Manufacturing Technology},
  volume={119},
  number={3},
  pages={1461--1478},
  year={2022},
  publisher={Springer}
}

@article{hategekimana2020iot,
  title={IoT Device security through dynamic hardware isolation with cloud-Based update},
  author={Hategekimana, Festus and Whitaker, Taylor JL and Pantho, Md Jubaer Hossain and Bobda, Christophe},
  journal={Journal of Systems Architecture},
  volume={109},
  pages={101827},
  year={2020},
  publisher={Elsevier}
}

@inproceedings{liu2017safernet,
  title={Safernet: Safe transportation routing in the era of internet of vehicles and mobile crowd sensing},
  author={Liu, Qun and Kumar, Suman and Mago, Vijay},
  booktitle={2017 14th IEEE Annual Consumer Communications \& Networking Conference (CCNC)},
  pages={299--304},
  year={2017},
  organization={IEEE}
}

@ARTICLE{10198233,

  author={Gupta, Maanak and Akiri, Charankumar and Aryal, Kshitiz and Parker, Eli and Praharaj, Lopamudra},

  journal={IEEE Access}, 

  title={From ChatGPT to ThreatGPT: Impact of Generative AI in Cybersecurity and Privacy}, 

  year={2023},

  volume={11},

  number={},

  pages={80218-80245},

 }

@ARTICLE{Duan2023,
  author={Duan, Qiang and Huang, Jun and Hu, Shijing and Deng, Ruijun and Lu, Zhihui and Yu, Shui},
  journal={IEEE Communications Surveys \& Tutorials}, 
  title={Combining Federated Learning and Edge Computing Toward Ubiquitous Intelligence in 6G Network: Challenges, Recent Advances, and Future Directions}, 
  year={2023},
  volume={25},
  number={4},
  pages={2892-2950},
  doi={10.1109/COMST.2023.3316615}}

@article{ALAA201748,
title = {A review of smart home applications based on Internet of Things},
journal = {Journal of Network and Computer Applications},
volume = {97},
pages = {48-65},
year = {2017},
issn = {1084-8045},
doi = {https://doi.org/10.1016/j.jnca.2017.08.017},
author = {Mussab Alaa and A.A. Zaidan and B.B. Zaidan and Mohammed Talal and M.L.M. Kiah}
}

@article{giordani2020toward,
  title={Toward 6G networks: Use cases and technologies},
  author={Giordani, Marco and Polese, Michele and Mezzavilla, Marco and Rangan, Sundeep and Zorzi, Michele},
  journal={IEEE Comm. Magazine},
  volume={58},
  number={3},
  pages={55--61},
  year={2020},
  publisher={IEEE}
}

@inproceedings{ogawa2019iot,
  title={IoT device virtualization for efficient resource utilization in smart city IoT platform},
  author={Ogawa, Keigo and Kanai, Kenji and Nakamura, Kenichi and Kanemitsu, Hidehiro and Katto, Jiro and Nakazato, Hidenori},
  booktitle={2019 IEEE International Conference on Pervasive Computing and Communications Workshops (PerCom Workshops)},
  pages={419--422},
  year={2019},
  organization={IEEE}
}

@article{sun2020machine,
  title={When machine learning meets privacy in 6G: A survey},
  author={Sun, Yuanyuan and Liu, Jiajia and Wang, Jiadai and Cao, Yurui and Kato, Nei},
  journal={IEEE Comm. Surveys \& Tutorials},
  volume={22},
  number={4},
  pages={2694--2724},
  year={2020},
  publisher={IEEE}
}

@article{alam2020survey,
  title={A survey of network virtualization techniques for Internet of Things using SDN and NFV},
  author={Alam, Iqbal and Sharif, Kashif and Li, Fan and Latif, Zohaib and Karim, Md Monjurul and Biswas, Sujit and Nour, Boubakr and Wang, Yu},
  journal={ACM Computing Surveys (CSUR)},
  volume={53},
  number={2},
  pages={1--40},
  year={2020},
  publisher={ACM New York, NY, USA}
}

@misc{garcia2015edge,
  title={Edge-centric computing: Vision and challenges},
  author={Garcia Lopez, Pedro and Montresor, Alberto and Epema, Dick and Datta, Anwitaman and Higashino, Teruo and Iamnitchi, Adriana and Barcellos, Marinho and Felber, Pascal and Riviere, Etienne},
  journal={ACM SIGCOMM Computer Communication Review},
  volume={45},
  number={5},
  pages={37--42},
  year={2015},
  publisher={ACM New York, NY, USA}
}

@article{rezaeibagha2021toward,
  title={Toward secure data computation and outsource for multi-user cloud-based IoT},
  author={Rezaeibagha, Fatemeh and Mu, Yi and Huang, Ke and Chen, Lanxiang and Zhang, Leyou},
  journal={IEEE Transactions on Cloud Computing},
  volume={11},
  number={1},
  pages={217--228},
  year={2021},
  publisher={IEEE}
}

@inproceedings{sasirekha2023synthesis,
  title={Synthesis of IoT Sensor Telemetry Data for Smart Home Edge-IDS Evaluation},
  author={Sasirekha, GVK and Bangari, Amulya and Rao, Madhav and Bapat, Jyotsna and Das, Debabrata},
  booktitle={2023 International Conference on Computer Science, Information Technology and Engineering (ICCoSITE)},
  pages={562--567},
  year={2023},
  organization={IEEE}
}

@inproceedings{srinivas2022deployment,
  title={Deployment Automation for Blockchain Enabled IoMT},
  author={Srinivas, Abburu Kalyan and Vikram, Deepa and Sharma, Suraj and others},
  booktitle={2022 OITS International Conference on Information Technology (OCIT)},
  pages={1--4},
  year={2022},
  organization={IEEE}
}

@inproceedings{vajagic2022smart,
  title={Smart Home IoT Network Diagnostics using Big Data Services},
  author={Vajagic, Nenad and Antic, Marija},
  booktitle={2022 30th Telecommunications Forum (TELFOR)},
  pages={1--4},
  year={2022},
  organization={IEEE}
}

@inproceedings{mehta2023big,
  title={Big Data Analytics Cloud based Smart IoT Healthcare Network},
  author={Mehta, Kapil and Gaur, Shivang and Maheshwari, Shubham and Chugh, Himani and anibhushan Kumar, M},
  booktitle={7th International Conference on Trends in Electronics and Informatics (ICOEI)},
  pages={437--443},
  year={2023},
}

@article{hahn2020efficient,
  title={Efficient IoT management with resilience to unauthorized access to cloud storage},
  author={Hahn, Changhee and Kim, Jongkil and Kwon, Hyunsoo and Hur, Junbeom},
  journal={IEEE transactions on cloud computing},
  volume={10},
  number={2},
  pages={1008--1020},
  year={2020},
  publisher={IEEE}
}

@inproceedings{mishra2023design,
  title={Design of a cloud-based security mechanism for Industry 4.0 communication},
  author={Mishra, Amit Kumar and Wazid, Mohammad},
  booktitle={2023 Third International Conference on Secure Cyber Computing and Communication (ICSCCC)},
  pages={337--343},
  year={2023},
  organization={IEEE}
}

@article{condon2022design,
  title={Design and implementation of a cloud-IoT-based home energy management system},
  author={Condon, Felipe and Mart{\'\i}nez, Jos{\'e} M and Eltamaly, Ali M and Kim, Young-Chon and Ahmed, Mohamed A},
  journal={Sensors},
  volume={23},
  number={1},
  pages={176},
  year={2022},
  publisher={MDPI}
}

@article{guo2021enabling,
  title={Enabling massive IoT toward 6G: A comprehensive survey},
  author={Guo, Fengxian and Yu, F Richard and Zhang, Heli and Li, Xi and Ji, Hong and Leung, Victor CM},
  journal={IEEE Internet of Things Journal},
  volume={8},
  number={15},
  pages={11891--11915},
  year={2021},
  publisher={IEEE}
}

@inproceedings{sai2023smart,
  title={Smart Home Messenger Notifications System using IoT},
  author={Sai, M Rama and Teja, K Krishna and Sasank, V Poorna and Kavitha, M and Aravinth, SS},
  booktitle={2023 Third International Conference on Artificial Intelligence and Smart Energy (ICAIS)},
  pages={87--92},
  year={2023},
  organization={IEEE}
}

@inproceedings{mladenova2023smart,
  title={Smart Home Based on IoT-Architecture and Practices},
  author={Mladenova, Tsvetelina and Cankov, Vladimir},
  booktitle={2023 5th International Congress on Human-Computer Interaction, Optimization and Robotic Applications (HORA)},
  pages={1--5},
  year={2023},
  organization={IEEE}
}

@article{adhikari20226g,
  title={6G-enabled ultra-reliable low-latency communication in edge networks},
  author={Adhikari, Mainak and Hazra, Abhishek},
  journal={IEEE Communications Standards Magazine},
  volume={6},
  number={1},
  pages={67--74},
  year={2022},
  publisher={IEEE}
}

@article{you2021towards,
  title={Towards 6G wireless communication networks: Vision, enabling technologies, and new paradigm shifts},
  author={You, Xiaohu and Wang, Cheng-Xiang and Huang, Jie and Gao, Xiqi and Zhang, Zaichen and Wang, Mao and Huang, Yongming and Zhang, Chuan and Jiang, Yanxiang and Wang, Jiaheng and others},
  journal={Science China Information Sciences},
  volume={64},
  pages={1--74},
  year={2021},
  publisher={Springer}
}

@inproceedings{antzoulis2022iot,
  title={IoT Security for Smart Home: Issues and Solutions},
  author={Antzoulis, Ioannis and Chowdhury, Md Minhaz and Latiff, Shadman},
  booktitle={2022 IEEE International Conference on Electro Information Technology (eIT)},
  pages={1--7},
  year={2022},
  organization={IEEE}
}

@article{prakash2022software,
  title={Software update practices on smart home IoT devices},
  author={Prakash, Vijay and Xie, Sicheng and Huang, Danny Yuxing},
  journal={arXiv preprint arXiv:2208.14367},
  year={2022}
}

@article{kolias2017ddos,
  title={DDoS in the IoT: Mirai and other botnets},
  author={Kolias, Constantinos and Kambourakis, Georgios and Stavrou, Angelos and Voas, Jeffrey},
  journal={Computer},
  volume={50},
  number={7},
  pages={80--84},
  year={2017},
  publisher={IEEE}
}

@article{mao2023security,
  title={Security and privacy on 6g network edge: A survey},
  author={Mao, Bomin and Liu, Jiajia and Wu, Yingying and Kato, Nei},
  journal={IEEE communications surveys \& tutorials},
  year={2023},
  publisher={IEEE}
}

@article{wang2022sustainable,
  title={Sustainable blockchain-based digital twin management architecture for IoT devices},
  author={Wang, Chenyu and Cai, Zhipeng and Li, Yingshu},
  journal={IEEE Internet of Things},
  volume={10},
  number={8},
  pages={6535--6548},
  year={2022},
  publisher={IEEE}
}
%

\end{document}